\documentstyle[12pt,epsfig,palatino]{article}
\def\beq{\begin{equation}}
\def\eeq{\end{equation}}
\def\bea{\begin{eqnarray}}
\def\eea{\end{eqnarray}}
\def\bq{\begin{quote}}
\def\eq{\end{quote}}

\def\bq{\begin{quote}}
\def\eq{\end{quote}}

\def\mpl{\ifmmode \overline M_{P}\else $\overline M_{P}$\fi}
\begin{document} 
\baselineskip 22pt
\vspace*{-1in} 
\renewcommand{\thefootnote}{\fnsymbol{footnote}} 
\begin{flushright} 
SINP/TNP/05-06~~~\\
TIFR/TH/05-17~~~~~~~\\
{\tt hep-ph/0506158}
\end{flushright} 
\vskip 65pt 
\begin{center} 
{\Large \bf NLO-QCD Corrections to Dilepton Production
in the Randall-Sundrum Model}\\
\vspace{8mm} 
{\bf 
Prakash Mathews$^{a}$
\footnote{prakash.mathews@saha.ac.in}, 
V. Ravindran$^b$
\footnote{ravindra@mri.ernet.in},   
K.~Sridhar$^c$
\footnote{sridhar@theory.tifr.res.in}
}\\ 
\end{center}
\vspace{10pt} 
\begin{flushleft}
{\it 
a) Theory Group, Saha Institute of Nuclear Physics, 1/AF Bidhan Nagar,\\ 
Kolkata 700 064, India.\\
 
b) Harish-Chandra Research Institute, 
 Chhatnag Road, Jhunsi, Allahabad, India.\\

c) Department of Theoretical Physics, 
Tata Institute of Fundamental Research,\\   
Homi Bhabha Road, Mumbai 400 005, India. \\

} 
\end{flushleft}
 
\vspace{10pt} 
\begin{center}
{\bf ABSTRACT} 
\end{center} 
\vskip12pt 
The dilepton production process at hadron colliders in the Randall-Sundrum
(RS) model is studied at next-to-leading order in QCD. The NLO-QCD corrections
have been computed for the virtual graviton exchange process in the RS model,
in addition to the usual $\gamma,\ Z$-mediated processes of standard 
Drell-Yan. $K$-factors for the cross-sections at the LHC and Tevatron for 
differential in the invariant mass, $Q$, and the rapidity, $Y$, of the lepton 
pair are presented.  We find the $K$-factors are large over substantial regions
of the phase space.
 
\vfill 
\clearpage 

\setcounter{page}{1} 
\pagestyle{plain}

\noindent 
In brane-world models, the four dimensional
universe is a dynamical hypersurface: a $D_3$-brane (or 3-brane) 
existing in a higher dimensional spacetime. In many such models, the 
Standard Model (SM) fields are localized on the brane and only gravity 
can propagate in the bulk. The scale of quantum gravity
can be lowered down from the Planck scale to the TeV scale in these
models \cite{string} making it exciting for high-energy physics
not only because these suggest fresh perspectives to the 
solution of the hierarchy problem but also because these models throw
open the possibility of the discovery of new physics at energies
accessible to collider experiments. In addition, these models
provide new frameworks for gauge symmetry and supersymmetry breaking
and suggest theoretical approaches to the cosmological constant
problem and dark-matter problem.

The simplest model seeking to address the gauge hierarchy problem was the
the ADD model proposed by
Arkani-Hamed, Dimopoulos and Dvali \cite{dimo}, where, starting from
a higher dimensional theory, an effective four-dimensional theory
at a scale $M_S \sim {\rm TeV}$ is obtained. This is done by compactifying 
the extra dimensions to magnitudes which are large compared to the Planck 
length \cite{revadd}. 

The main problem that one faces within the ADD model is the 
reappearance of disparate scales $viz.$, the string scale and
the inverse of the compactification radius.
It was an attempt to avoid this problem that led to the formulation
of the Randall-Sundrum (RS) model \cite{rs}.
In the RS model 
the single extra dimension $\phi$ is compactified on a ${\bf
S}^1/{\bf Z}^2$ orbifold with a radius $R_c$ which is 
somewhat larger than the Planck length. 
Two 3-branes, the Planck brane and the TeV brane, are located
at the orbifold fixed points $\phi=0,\ \pi$, with the  
SM fields localised on the TeV brane.
The five-dimensional metric, which is {\it non-factorisable} or $warped$ 
is of the form
\begin{equation}
ds^2 = e^{-{\cal K}R_c\phi}\eta_{\mu\nu}dx^{\mu}dx^{\nu}~+~R_c^2d\phi^2  .
\label{eq1}
\end{equation} 
The exponential warp factor $e^{-{\cal K}R_c\phi}$ serves as a conformal factor 
for fields localised on the brane. 
Thus the huge ratio
$\frac{M_P}{M_{EW}} \sim 10^{15}$ can be generated by the exponent
${\cal K}R_c$ which needs to be only of
${\cal O}(10)$ thereby providing a way of avoiding the hierarchy problem. 
There remains the problems of stabilising 
$R_c$ against quantum fluctuations
but this can be done by introducing an extra scalar field in
the bulk \cite{gold, csaki}. 

The tower of massive Kaluza-Klein (KK) excitations of the graviton, 
$h^{(\vec{n})}_{\mu\nu}$, interact with the SM particles by:  
\begin{eqnarray} 
{\cal L}_{int} & \sim & -{1\over M_P}
T^{\mu\nu}(x) h^{(0)}_{\mu\nu}(x) 
-{e^{\pi {\cal K} R_c} \over M_P} \sum_1^{\infty}
T^{\mu\nu}(x) h^{(n)}_{\mu\nu}(x) \ . 
\label{eq2}
\end{eqnarray} 
$T^{\mu\nu}$ is the
symmetric energy-momentum tensor for the particles on the
3-brane. The masses of the $h^{(\vec{n})}_{\mu\nu}$ are given by 
\begin{eqnarray} 
M_n & = & x_n {\cal K} ~e^{-\pi {\cal K} R_c} \ ,
\label{eq3}
\end{eqnarray} 
where the $x_n$ are the zeros of the Bessel function $J_1(x)$. 
The zero-mode couples weakly and decouples
but the couplings of the
massive RS gravitons are enhanced by the exponential $e^{\pi {\cal K}
R_c}$ leading to interactions of electroweak strength. 
Consequently, except for the overall warp factor in the RS case,
the Feynman rules in the RS model are the same as those for the ADD case
\cite{grw, hlz}. 

The basic parameters of the RS model are 
\begin{eqnarray}
m_0 & = & {\cal K} e^{-\pi {\cal K} R_c} \ , \nonumber \\
c_0 & = & {\cal K}/M_P \ , 
\label{eq4}
\end{eqnarray}
where $m_0$ is a scale of the dimension of mass and sets the scale for 
the masses of the KK excitations, and $c_0$ is
an effective coupling. The interaction of massive KK gravitons with
the SM fields can be written as 
\begin{equation} 
{\cal L}_{int}  \sim 
- {c_0 \over m_0} \sum_n^{\infty} T^{\mu\nu}(x)
 h^{(n)}_{\mu\nu}(x) \ . 
\label{eq5}
\end{equation} 
Since ${\cal K}$ is related to the curvature of the fifth dimension
we need to restrict it to small enough values to avoid effects
of strong curvature. On the other hand ${\cal K}$ should not
be too small compared to $M_P$ because that would reintroduce
a hierarchy. These considerations suggest $0.01 \le c_0 \le 0.1$. 
For our analysis we choose to work with the RS parameters $c_0$ and 
$M_1$ the first excited mode of the graviton rather then $m_0$.  

The decoupling of the graviton zero-mode and the existence of a mass gap
in the spectrum of KK gravitons imply that it is only the resonant
production and decay of the heavier KK modes or the virtual effects
of the KK modes that one can hope to detect in collider experiments.
The phenomenology of resonant production of the KK excitations and the 
virtual effects have already been studied in processes like dilepton 
production \cite{dhr}, diphoton production \cite{sridhar}, $t \bar t$ 
production at hadron colliders \cite{lmrs}, $\tau$-production at a 
linear collider \cite{namit} and pair production of KK modes in
$e^+e^-$ and hadron hadron colliders \cite{jp}.  The sensitivity of 
the CMS experiment to the resonant production of RS graviton KK modes 
has been studied for electron pair production \cite{cms}.  Recently 
D\O\ has reported the first direct search for RS graviton KK modes 
using dielectron, dimuon, and diphoton events \cite{d0}.

In an earlier work we had presented NLO-QCD corrections for $e^+ e^- \to$ 
hadrons \cite{us} and dilepton pair production at hadron colliders \cite{us1}
in the ADD model. These results for the dilepton pair production case are 
extended to the RS model, in this paper. We 
note that it is the same virtual graviton exchange process that contributes 
to dilepton production in both the ADD and RS models. The leading order 
process being the same, the QCD
corrections are also not model-dependent. However, as explained above,
the differences between the two models arise because of the difference
in the summation over the tower of KK gravitons and also in the overall
factors. Consequently, the relative weight of the subprocess cross-section
due to graviton exchange vis-a-vis the SM subprocess will be different
in the two models. This results in different $K$-factors in the ADD
and RS models and the dependence of the $K$-factors on the kinematic 
variables are also different. In this letter, we present the results
for dilepton production at the LHC and Tevatron in the RS model.

The process we are interested in is where two hadrons $P_1,P_2$ scatter 
and give rise to leptonic final states,
say $\mu^+,\mu^-$  
\begin{eqnarray}
P_1(p_1)+P_2(p_2) \rightarrow \mu^+(l_1)+\mu^-(l_2)+X(P_X)  \ ,
\label{eq6}
\end{eqnarray}
where $p_1,p_2$ are the momenta of incoming hadrons $P_1$ and $P_2$
respectively and $\mu^-,\mu^+$ are the outgoing leptons which have
the momenta $l_1,l_2$. 
The final inclusive hadronic state is denoted by $X$  
and carries the momentum $P_X$.
The hadronic cross section
can be expressed in terms of partonic cross sections
convoluted with appropriate parton distribution functions as follows
\begin{eqnarray}
2 S~{d \sigma^{P_1 P_2} \over d Q^2}\left(\tau,Q^2\right)
&=&\sum_{ab={q,\overline q,g}} \int_0^1 dx_1
\int_0^1 dx_2~ f_a^{P_1}(x_1) ~
f_b^{P_2}(x_2)
\nonumber\\[2ex] &&
\times \int_0^1 dz \,\, 2 \hat s ~
{d \hat \sigma^{ab} \over d Q^2}\left(z,Q^2\right)
\delta(\tau-z x_1 x_2)\ .
\label{eq7}
\end{eqnarray}
The scaling variables are defined 
by $k_1 =x_1 p_1,k_2=x_2 p_2$ where $k_1,k_2$ are the momenta of
incoming partons.  
\begin{eqnarray}
(p_1+p_2)^2 &\equiv& S, \quad \quad \quad 
(k_1+k_2)^2 \equiv \hat s, \quad \quad \quad (l_1+l_2)^2=q.q \equiv Q^2,
\nonumber\\[2ex]
\tau&=&{Q^2 \over S} \ , 
\quad \quad \quad 
z={Q^2 \over \hat s } \ , 
\quad \quad \quad \tau=x_1 x_2 z \ .
\label{eq8}
\end{eqnarray}
The partonic cross section for the process
$a(k_1)+b(k_2) \rightarrow j(-q)+\displaystyle \sum_i^m X_i(-p_i)$ 
is given by
\begin{eqnarray}
2 \hat s ~ {d \hat \sigma^{ab} \over d Q^2} &=&
{1 \over 2 \pi} \sum_{jj'=\gamma,Z,G}
\int dPS_{m+1}~  |M^{ab \rightarrow jj'}|^2\cdot P_j(q)\cdot P^*_{j'}(q)\cdot 
{\cal L}^{jj' \rightarrow l^+l^-}(q) \ .
\label{eq9}
\end{eqnarray}
In the above equation, the sum over Lorentz indices between 
matrix element squared and the propagators is implicit through
a symbol ``dot product".
The $m+1$ body phase space is defined as
\begin{eqnarray}
\int dPS_{m+1}&=&\int \prod_i^m \Bigg({d^n p_i \over (2\pi)^n} 
2 \pi \delta^+(p_i^2)\Bigg) {d^nq\over (2\pi)^n} 2 \pi \delta^+(q^2-Q^2)
\nonumber\\[2ex]&&
\times (2 \pi)^n \delta^{(n)}(k_1+k_2+q+\sum_i^m p_i)\ , 
\label{eq10}
\end{eqnarray}
where $n$ is the space-time dimension.
The propagators are
\begin{eqnarray}
P_{\gamma}(q)&=&-{i \over Q^2} g_{\mu \nu}\ ,
\label{eq11}
\\ [2ex]
P_Z(q) &=& ~-{i \over 
(Q^2 -M_Z^2 - i M_Z \Gamma_Z)} g_{\mu\nu} \ ,
\label{eq12}
\\ [2ex]
P_{G}(q)&=&{\cal D}(Q^2) B_{\mu \nu \lambda \rho} (q) \ ,
\label{eq13}
\end{eqnarray}
where 
\begin{eqnarray}
B_{\mu \nu \rho \sigma}(q)&=&
\Bigg(g_{\mu \rho} - {q_\mu q_\rho \over {M_n}^2} \Bigg)
\Bigg(g_{\nu \sigma} - {q_\nu q_\sigma \over {M_n}^2} \Bigg)
+\Bigg(g_{\mu \sigma} - {q_\mu q_\sigma \over {M_n}^2} \Bigg)
\Bigg(g_{\nu \rho} - {q_\nu q_\rho \over {M_n}^2} \Bigg)
\nonumber\\[2ex]
&& -{2 \over n-1} \Bigg(g_{\mu \nu} - {q_\mu q_\nu \over {M_n}^2} \Bigg)
\Bigg(g_{\rho \sigma} - {q_\rho q_\sigma \over {M_n}^2} \Bigg) \ .
\label{eq14}
\end{eqnarray}
The function ${\cal D}(Q^2)$ in the graviton propagator Eq.~(\ref{eq13}), 
results from summing over the KK modes, given by 
\begin{eqnarray}
{\cal D}(Q^2) &=& \sum_{n=1}^\infty \frac{1}{Q^2 - M_n^2 + i M_n \Gamma_n} 
\equiv {\lambda \over m_0^2} \ ,
\label{eq15}
\end{eqnarray}
where $M_n$ are the masses of the individual resonances and the $\Gamma_n$ 
are the corresponding widths.  The graviton widths are obtained by calculating 
their decays into final states involving SM particles.  $\lambda$ is defined 
as 
\begin{eqnarray}
\lambda (x_s) & = & \sum_{n=1}^\infty 
\frac{x_s^2 -x_n^2 -i \frac{\Gamma_n}{m_0} x_n}
     {x_s^2 -x_n^2 +  \frac{\Gamma_n}{m_0} x_n} \ ,
\label{eq16}
\end{eqnarray}
where $x_s=Q/m_0$.  We have to 
sum over all the resonances to get the value of $\lambda(x_s)$. This is done 
numerically and for a given value of $x_s$, we retain all resonances which 
contribute with a significance greater than one per mil, and treat the 
remaining KK modes as virtual particles (in which case the sum can be done 
analytically).

We now present the distributions in the invariant lepton pair mass, $Q$, 
and the rapidity of the lepton pair, $Y$ at the LHC ($\sqrt{S}=14$ 
TeV) and Tevatron ($\sqrt{S}=1.96$ TeV).  From these distributions 
the effects of the NLO-QCD corrections can be clearly discerned.  For 
the parton density sets we adopt in leading order (LO) MRST 2001 LO 
($\Lambda=0.1670~{\rm GeV}$) and in next-to-leading order the MRST 2001 
NLO ($\Lambda=0.2390$ GeV). For LHC we choose the kinematic ranges $300~
{\rm GeV}<Q<3000~{\rm GeV}$ and $|Y|<2.2$ at $Q=1.5$ TeV.  For Tevatron 
$300~{\rm GeV}<Q<1000$ GeV and $|Y| <0.9$ at $Q=300$ GeV.  The renormalisation
scale is taken to be same as the factorization scale $\mu_F$ and $\mu_F$ 
si chosen to be $\mu_F=Q$.

The cross-section $d\sigma/dQ$ as a function of $Q$ to NLO is presented in 
Fig.~1a for LHC.  For the figure, we have chosen the representative values 
of the RS model parameters: $M_1 =1.5$ TeV the first RS resonance mass and 
the coupling constant $c_0= 0.01$.  The width of the resonance is related 
to $c_0$ and hence a smaller $c_0$ corresponds to a narrow resonance.  The 
subsequent resonance are determined by $m_0$ and $x_n$.  To LO the dilepton 
case has been presented in \cite{dhr}.  To see the effect of the NLO effect 
we study the $K$-factor for the $Q$ and $Y$ distribution.  

The $K$-factor for the invariant lepton pair mass distribution defined by 
\begin{eqnarray}
K^I=\Bigg[ {d \sigma_{LO}^I(Q) \over dQ} \Bigg]^{-1}
       \Bigg[ {d \sigma_{NLO}^I(Q) \over dQ} \Bigg] \ ,
\label{eq17}
\end{eqnarray}
where $I=SM$, $I=SM+GR$ for both SM and gravity combined and $I=GR$ for 
only gravity.  It is possible to define $K^{GR}$ for the invariant 
lepton pair mass distribution, as there is no interference with SM 
\cite{us1}.  The results are presented in Fig.~1b. The parameters chosen 
are the same as in Fig.~1a.  In order to understand the 
behaviour of $K$-factor of the model involving both SM and gravity, it is 
useful to express it as
\begin{eqnarray}
K^{(SM+GR)}(Q) = \frac{K^{SM} + K^{GR} K^{(0)}}{1+K^{(0)}} \ ,
\label{eq18}
\end{eqnarray} 
where we have introduced a quantity $K^{(0)}$, defined as the ratio of the 
LO distribution of gravity to SM, given by 
\begin{eqnarray}
K^{(0)}(Q)=\Bigg[ {d \sigma_{LO}^{SM}(Q) \over dQ} \Bigg]^{-1}
       \Bigg[ {d \sigma_{LO}^{GR}(Q) \over dQ} \Bigg] \ .
\label{eq19}
\end{eqnarray}
The behaviour of $K^{(0)}(Q)$ is governed by competing couplings constants 
of SM and gravity and the parton fluxes.  In the RS case the gravity 
contribution is significant in the resonance region, (see Fig.1a).  In 
the off resonance region the $K$-factor is hence purely $K^{SM}$.  In the 
resonance region where the gravity effect dominates the $K^{(SM+GR)}$ factor
shifts to the $K^{GR}$ value (see Fig.~1b).  This behaviour of the $K$-factor 
of the RS case is very distinct from the corresponding case we presented in 
the ADD case \cite{us1}.  To incorporate the NLO effects for an 
appropriate distribution one needs to take into account the behaviour of the 
$K$-factor accordingly.  For $M_1=300$ GeV the $K$-factor is about $1.5$
in the resonance region.   This is due to the fact that at loq $Q$ ($Q=300$ 
GeV) the gluon flux becomes dominant at Tevatron.  The behaviour of $K^{GR}$ 
is the same as in the ADD case \cite{us1}.  

In Fig.~1c, we have plotted the scale variations of the $Q$
distribution for both LO and NLO cross sections. We define $R^I$ for 
the invariant lepton mass distribution as 
\begin{eqnarray}
R_{LO}^I&=&\Bigg[ {d \sigma_{LO}^I(Q,\mu=\mu_0) \over dQ} \Bigg]^{-1}
       \Bigg[ {d \sigma_{LO}^I(Q,\mu) \over dQ } \Bigg] \ ,
\nonumber\\[2ex]
R_{NLO}^I&=&\Bigg[ {d \sigma_{NLO}^I(Q,\mu=\mu_0) \over dQ} \Bigg]^{-1}
       \Bigg[ {d \sigma_{NLO}^I(Q,\mu) \over dQ} \Bigg] \ ,
\label{eq20}
\end{eqnarray}
where $\mu_0$ is a fixed scale which is chosen to be $\mu_0=1.5$ TeV for
LHC.  As can be seen from the figure, the inclusion of the NLO corrections
stabilises the cross-section with respect to the scale $\mu$.  Here 
we have chosen $\mu_0=1.5$ TeV, ie. the first resonance region.  The 
scale variation is driven by the gravity part as its the dominant 
contribution.

In Fig.~2a the double differential cross section $d^2\sigma/dQdY$
is displayed for rapidity region $\vert Y \vert \le 2.2$ for a $Q$ value of 
1.5 TeV.  To plot this distribution the $Q$ value is chosen such that it 
lies at the first resonance, where the gravity effect dominates.  Hence the 
dominant contribution is purely gravity.  The RS model parameters remain the 
same as before.  The $K$ factor as a function of $Y$ is plotted for a choice 
of $Q$ in the resonance region where the dominant contribution to 
$K^{(SM+GR)}$ factor comes from the gravity part Fig.~2b.  The $R$ ratio using 
the $Y$ distribution is plotted in Fig.~2c for the central region of rapidity 
and for a $Q$ value of 1.5 TeV.  In this region the scale variation is also
dictated by the gravity contribution.  

The corresponding analysis for the Tevatron is done for the $Q$ range
$300 < Q < 1000$ GeV and for the RS parameter $M_1=300$ GeV and the 
coupling $c_0=0.01$.  At low $Q$ the gravity effects of the RS model is 
dominant in the resonance and off the resonance region the effect is 
negligible.  As $Q$ increases the effect of gravity starts to become
comparable to the SM contribution as is seen towards the third resonance 
in Fig.~3a.  In Fig.~3b we have plotted the $K$-factor for $Q$ 
distribution at the Tevatron.  Using Eq.~(\ref{eq18}) we can understand 
the behaviour of $K^{SM+GR}$.  As expected the behaviour of $K^{GR}$ is 
same as the ADD case \cite{us1}.  The double differential cross section 
$d^2\sigma /dQdY$ for $Q=300$ GeV is plotted as a function of rapidity $Y$.  
In the resonance region the dominant contribution is from the RS.
In contrast for ADD \cite{us1} only at large $Q$ the gravity effects became 
comparable to the SM at Tevatron.  The corresponding $K$-factor is plotted 
in Fig.~4b.  Scale variation for the $Q$ and double differential $dQdY$
is given in Fig.~3c and Fig.~4c respectively.

In a recent analysis by D\O\ \cite{d0}, the LO cross section was scaled by a 
constant $K$-factor of 1.34 to account for the NLO effect for the RS case.  
This does not yield a realistic picture as can be seen from Fig.~3b.  In
the RS case due to the resonant production, the $K$-factor is very different 
from the ADD case reported earlier \cite{us1}.    

In summary, we have presented the results for the cross-section for
dilepton production in the Randall-Sundrum model at the LHC and Tevatron. 
The large incident gluon flux at the LHC makes the NLO QCD corrections
very important. Moreover, when the NLO corrections are taken into
account the cross-sections are stabilised with respect to scale
variations. In order to derive robust bounds on the RS model at
the LHC using the dilepton production process, the inclusion
of the NLO QCD corrections in the cross-section is crucial.

\vspace{20pt}
\noindent
{\it Acknowledgments:} PM thanks S.~Moretti for useful discussion.  VR would 
like to thank Prof.~W.~L.~van Neerven for discussion.  The work of PM and KS 
is part of a project (IFCPAR Project No. 2904-2) on `Brane-World Phenomenology'
supported by the Indo-French Centre for the Promotion of Advanced Research, 
New Delhi, India. PM would also like to thank IPPP, Durham for warm hospitality
where part of this work was done.  PM and VR thank S.~Raychaudhuri for 
providing the code that evaluates the RS KK mode sum in the propagator.   

\eject
\centerline{\large \bf Figure Caption} 
\vspace{.5cm}

\noindent
Figure 1. (a) The cross section is plotted as a function of invariant mass 
$Q$ of the lepton pair for $M_1=1.5$ TeV at LHC.  (b) The corresponding 
$K$-factor for $Q$ distribution SM, gravity and SM plus gravity. (c) Scale 
variation of the cross section at LO and NLO as defined in Eq.~(\ref{eq19}) 
for $Q=1.5$ TeV. \\[1ex]

\noindent
Figure 2. (a) The double differential cross section $d^2 \sigma/dQ dY$ is 
plotted as a function of rapidity $Y$ for $Q=1.5$ TeV at LHC. 
(b) The K-factor for the distribution in (a) is plotted for the rapidity 
range.  (c) The scale variation of the ratio R is plotted as a function 
of $\mu/\mu_0$ for $Y=0$. \\[1ex]

\noindent
Figure 3. (a) The cross section is plotted as a function of invariant mass
$Q$ of the lepton pair for $M_1=300$ GeV and $c_0=0.01$ at the Tevatron. (b)
The $K$-factor for $Q$ distribution for the same RS parameters in (a) is
plotted. (c) The variation of the cross section with respect to the scale. 
\\[1ex]

\noindent
Figure 4. (a) The double differential cross section $d^2 \sigma/dQ dY$ is
plotted as a function of rapidity $Y$ for $Q=300$ GeV at the Tevatron for
the RS parameters $M_1=300$ GeV and $c_0=0.01$.  (b) The $K$-factor for the 
distribution in (a) is plotted for the rapidity range.  (c) The scale 
variation of the ratio R is plotted as a function of $\mu/\mu_0$ for the
central rapidity region $Y=0$.

\eject

\begin{figure}[htb]
\vspace{1mm}
\centerline{\epsfig{file=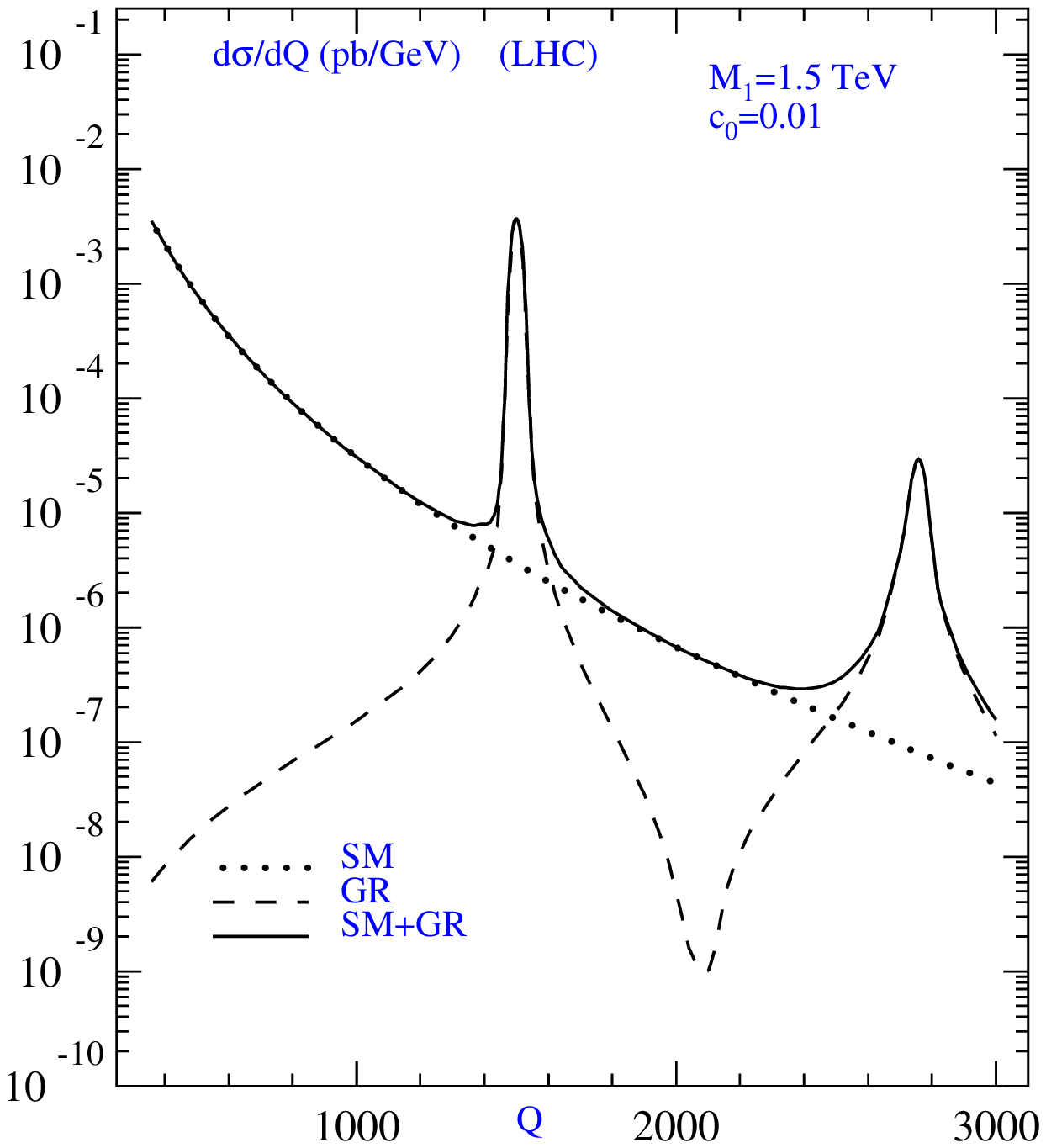,width=15cm,height=16cm,angle=0}}
\vspace{5mm}
\centerline{\bf Fig.~1a}
\end{figure}

\eject

\begin{figure}[htb]
\vspace{1mm}
\centerline{\epsfig{file=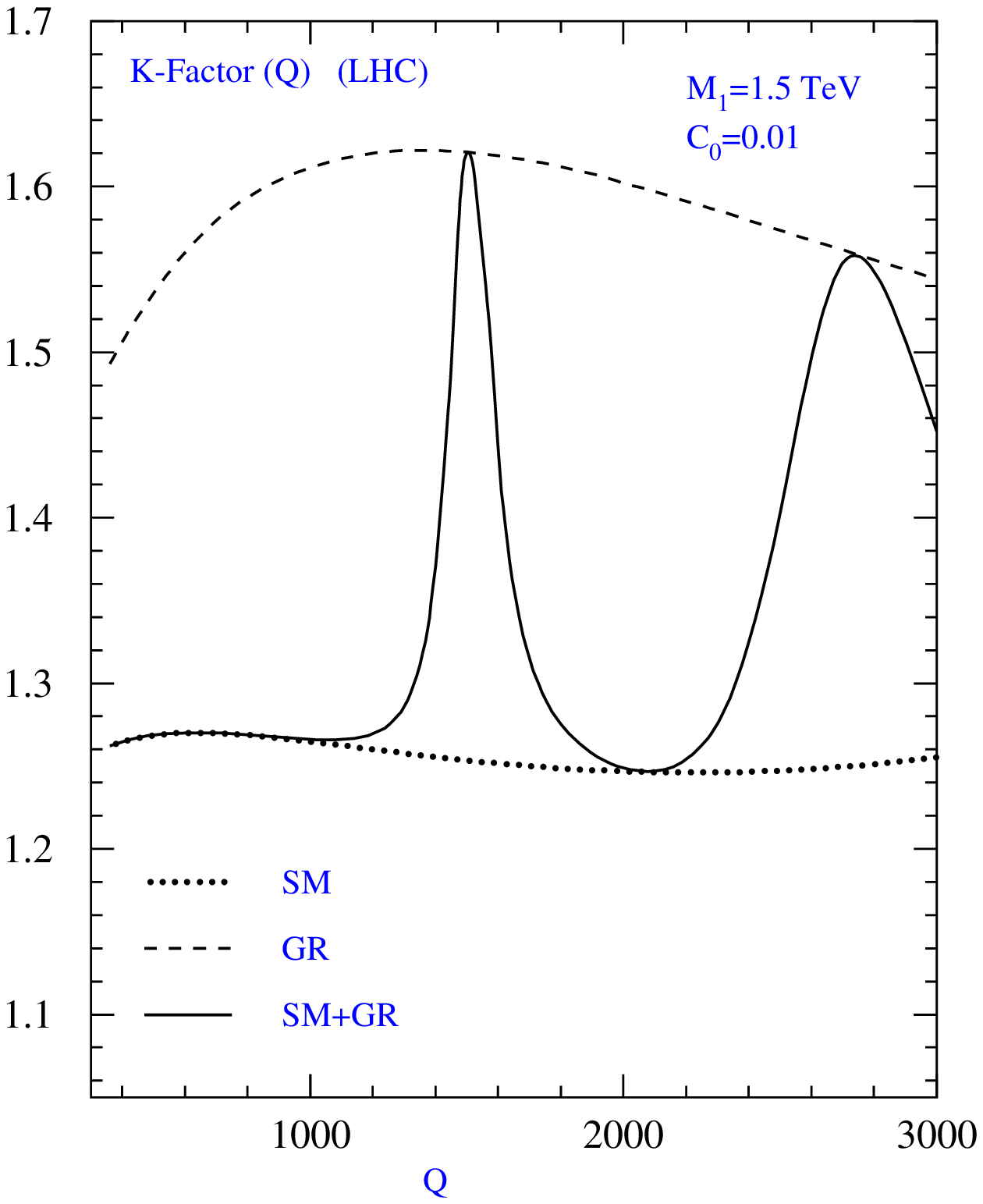,width=15cm,height=16cm,angle=0}}
\vspace{5mm}
\centerline{\bf Fig.~1b}
\end{figure}

\eject

\begin{figure}[htb]
\vspace{1mm}
\centerline{\epsfig{file=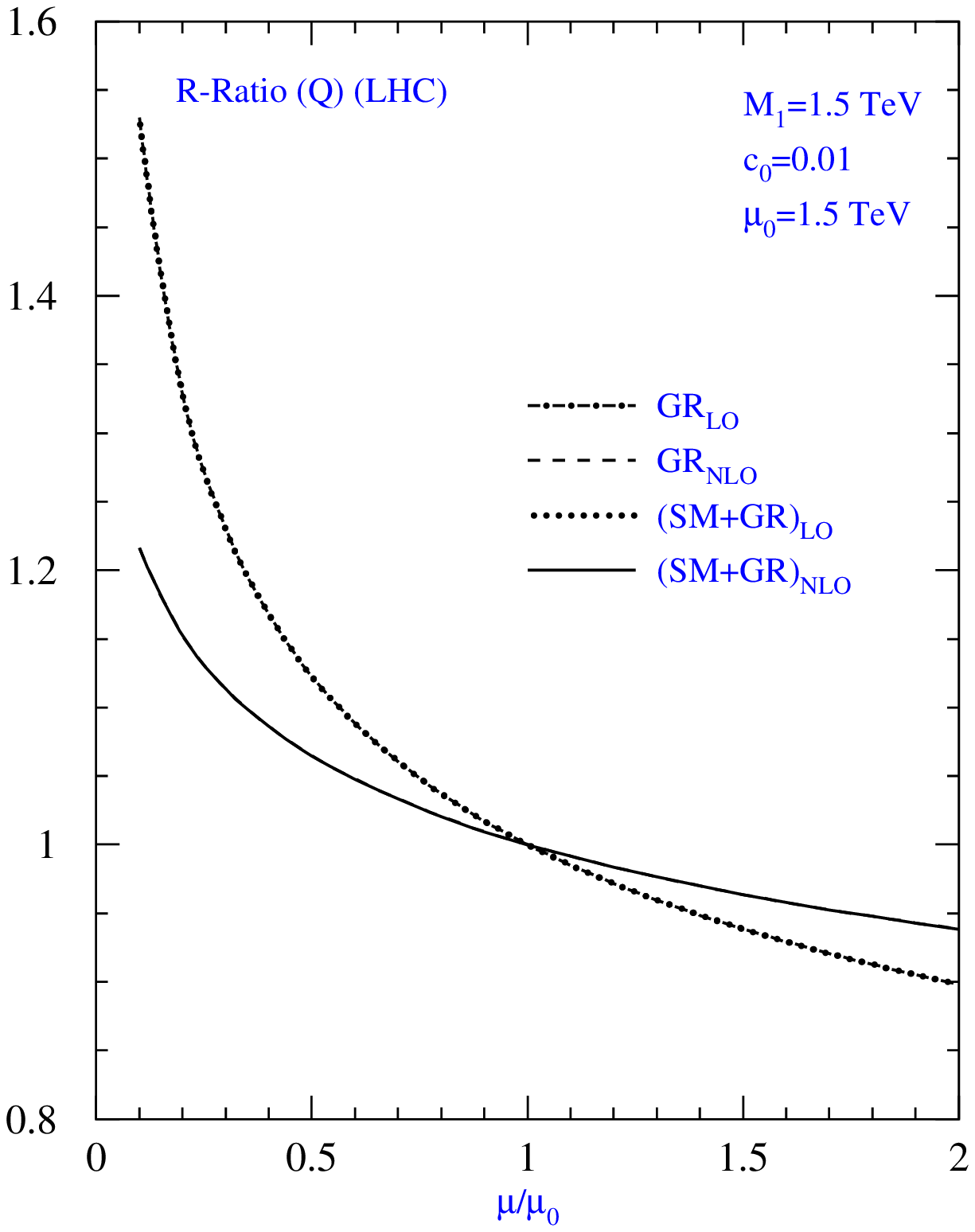,width=15cm,height=16cm,angle=0}}
\vspace{5mm}
\centerline{\bf Fig.~1c}
\end{figure}

\eject

\begin{figure}[htb]
\vspace{1mm}
\centerline{\epsfig{file=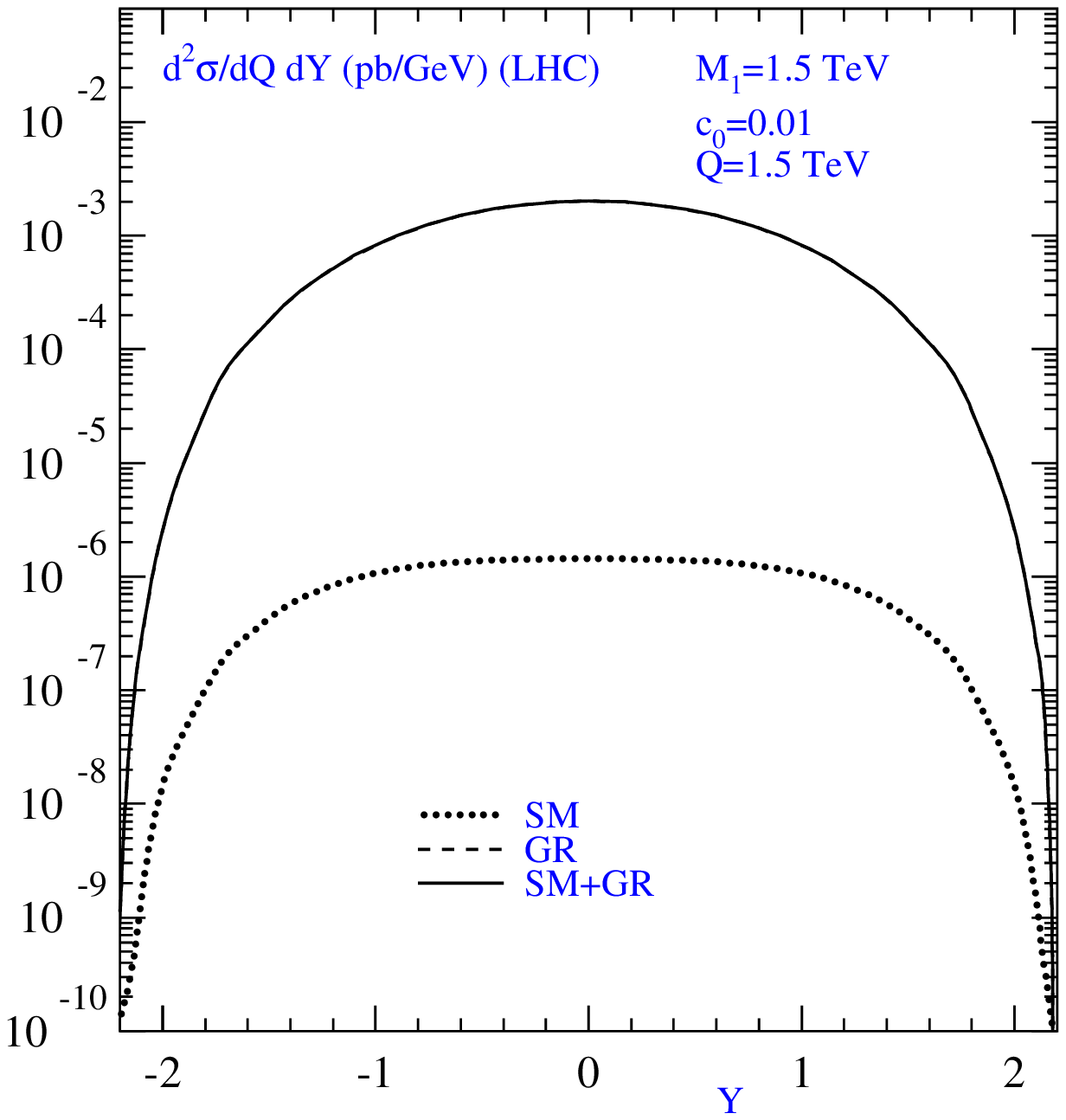,width=15cm,height=16cm,angle=0}}
\vspace{5mm}
\centerline{\bf Fig.~2a}
\end{figure}

\eject

\begin{figure}[htb]
\vspace{1mm}
\centerline{\epsfig{file=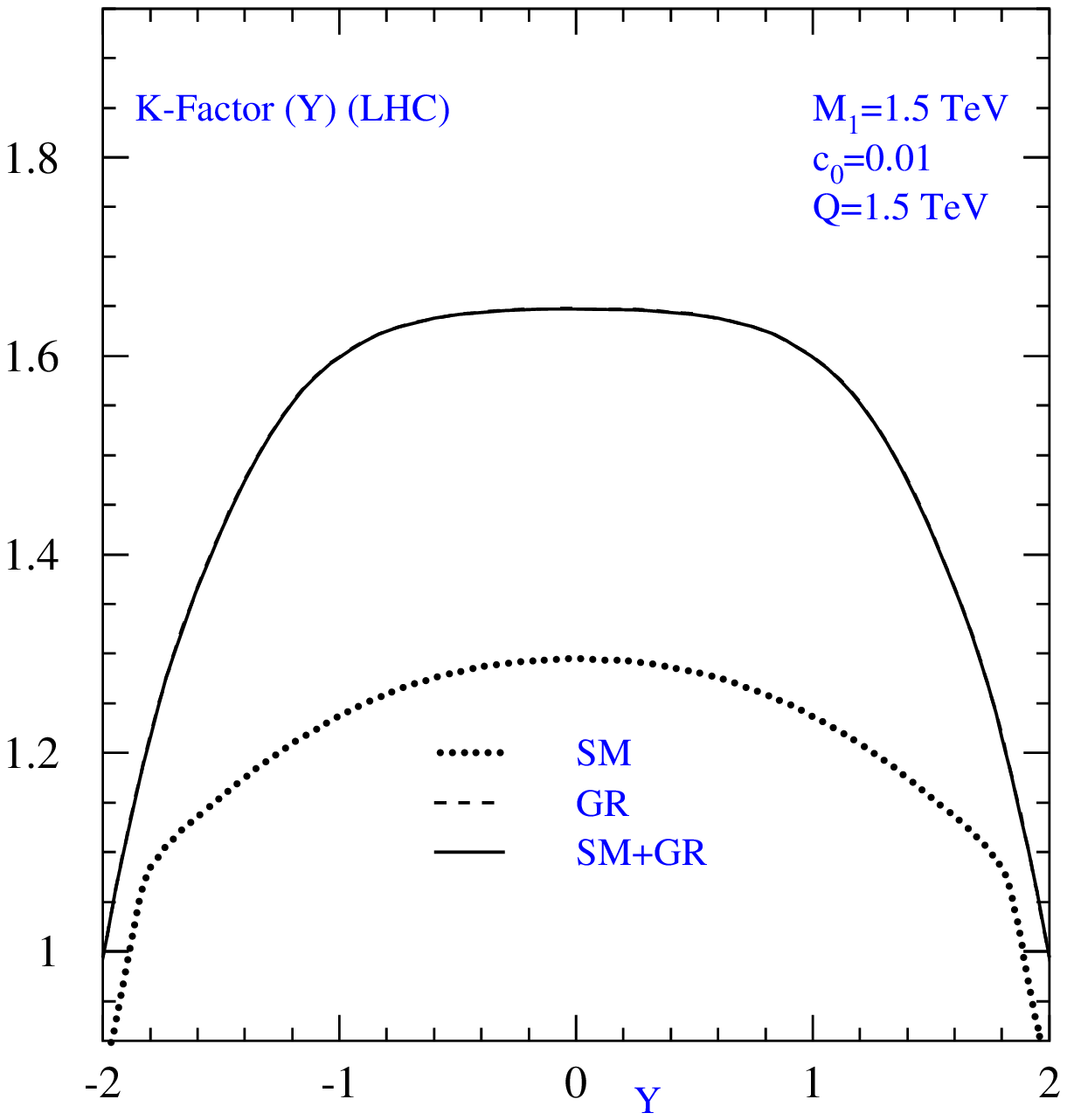,width=15cm,height=16cm,angle=0}}
\vspace{5mm}
\centerline{\bf Fig.~2b}
\end{figure}

\eject

\begin{figure}[htb]
\vspace{1mm}
\centerline{\epsfig{file=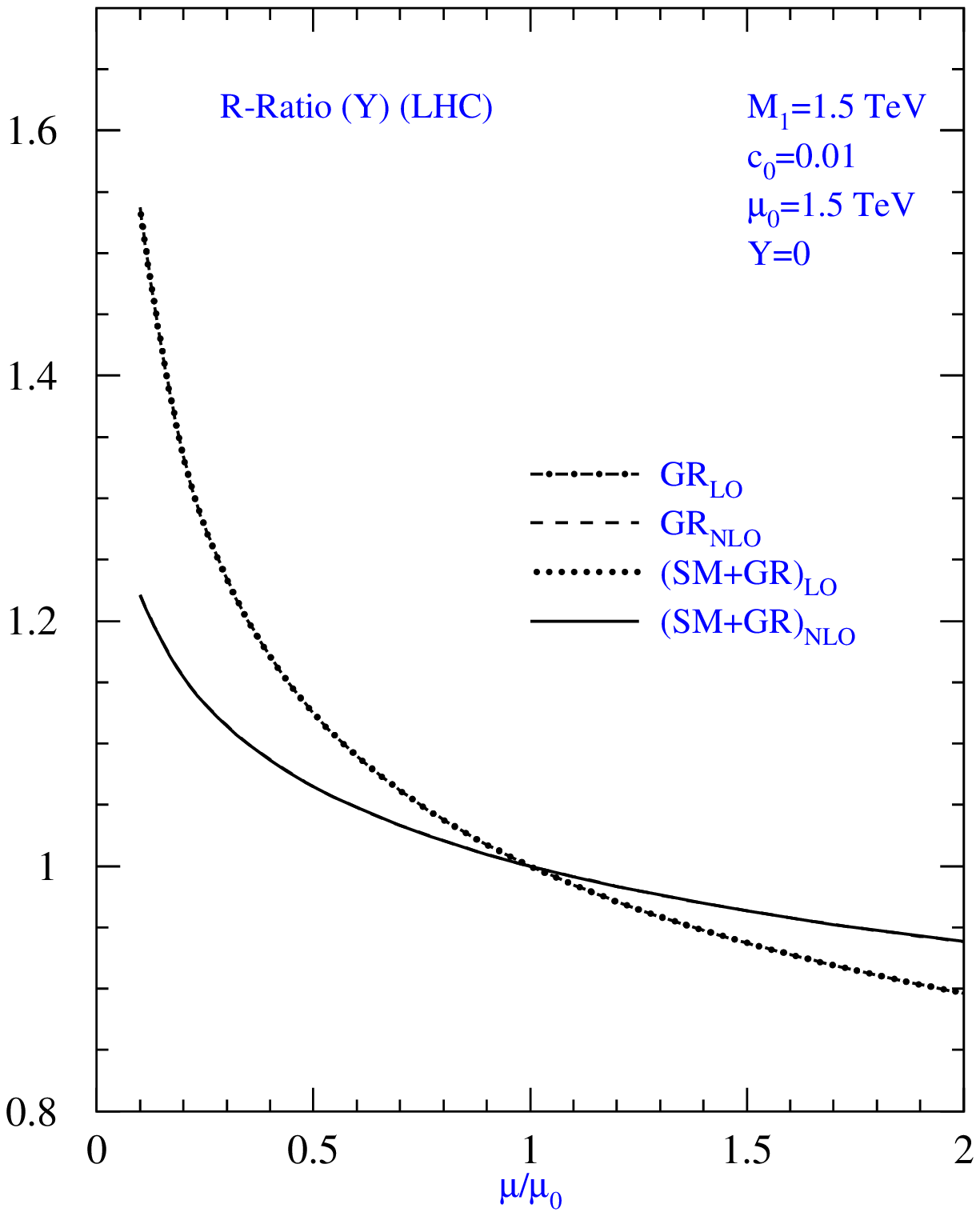,width=15cm,height=16cm,angle=0}}
\vspace{5mm}
\centerline{\bf Fig.~2c}
\end{figure}

\eject

\begin{figure}[htb]
\vspace{1mm}
\centerline{\epsfig{file=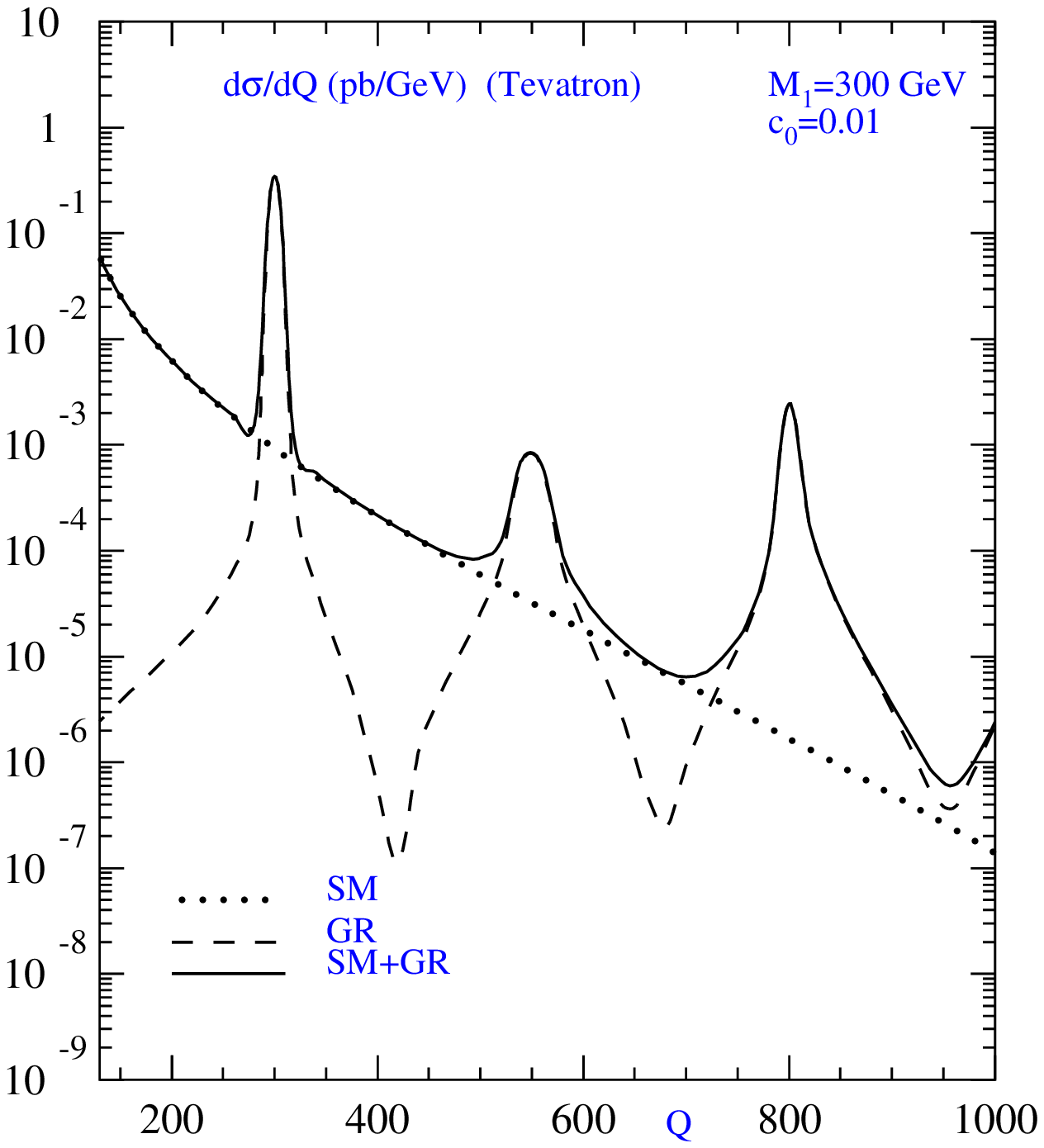,width=15cm,height=16cm,angle=0}}
\vspace{5mm}
\centerline{\bf Fig.~3a}
\end{figure}

\eject

\begin{figure}[htb]
\vspace{1mm}
\centerline{\epsfig{file=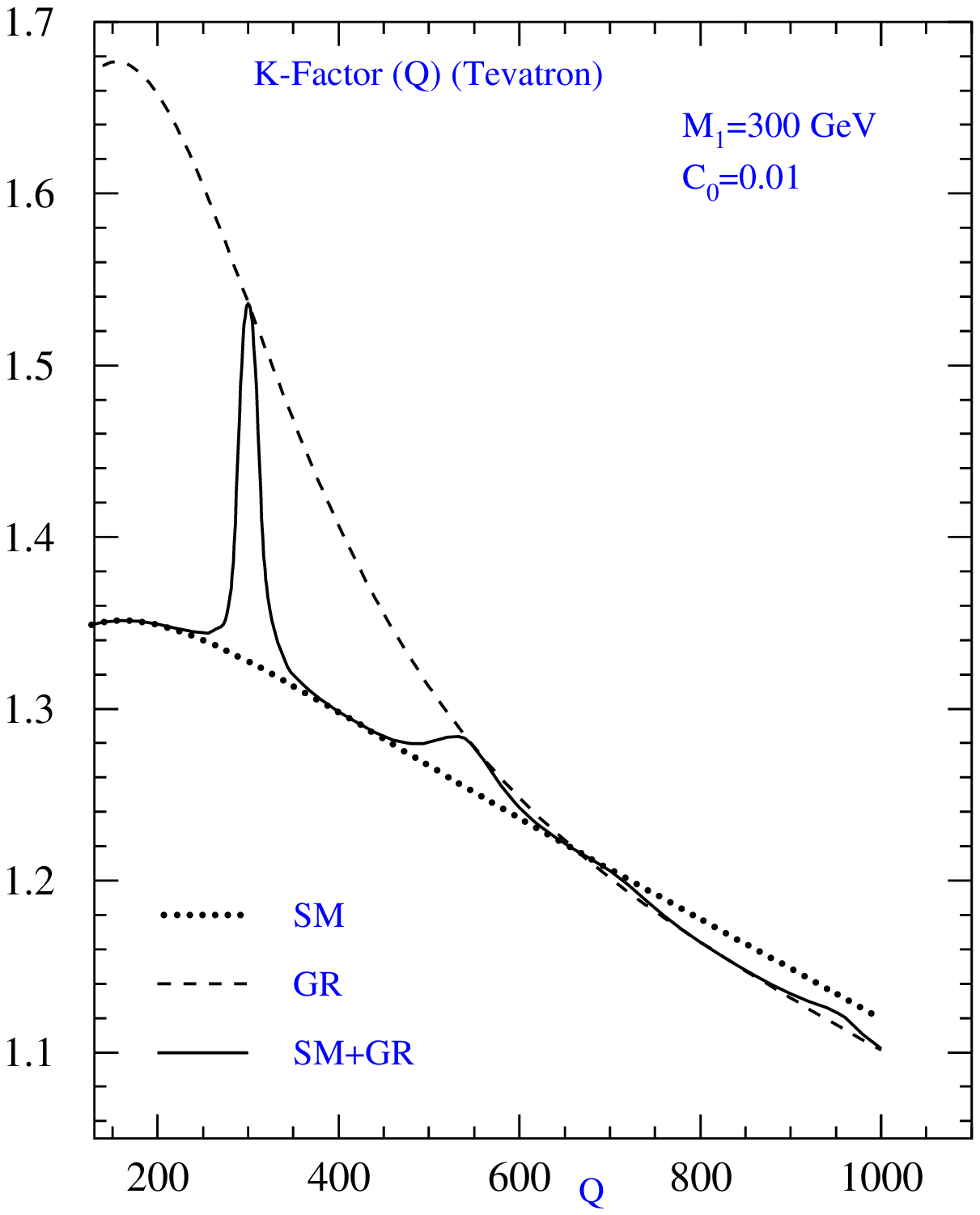,width=15cm,height=16cm,angle=0}}
\vspace{5mm}
\centerline{\bf Fig.~3b}
\end{figure}

\eject

\begin{figure}[htb]
\vspace{1mm}
\centerline{\epsfig{file=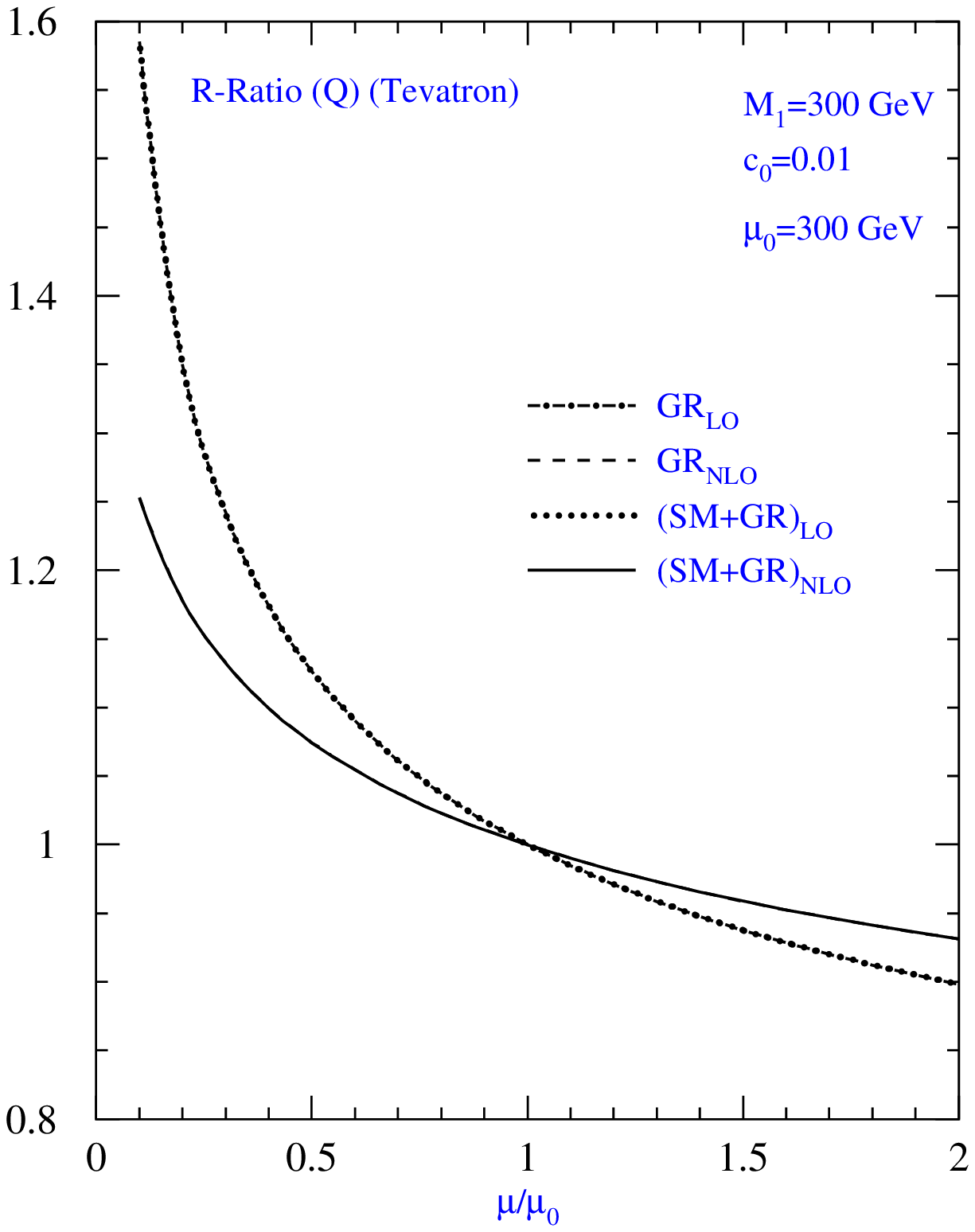,width=15cm,height=16cm,angle=0}}
\vspace{5mm}
\centerline{\bf Fig.~3c}
\end{figure}

\eject

\begin{figure}[htb]
\vspace{1mm}
\centerline{\epsfig{file=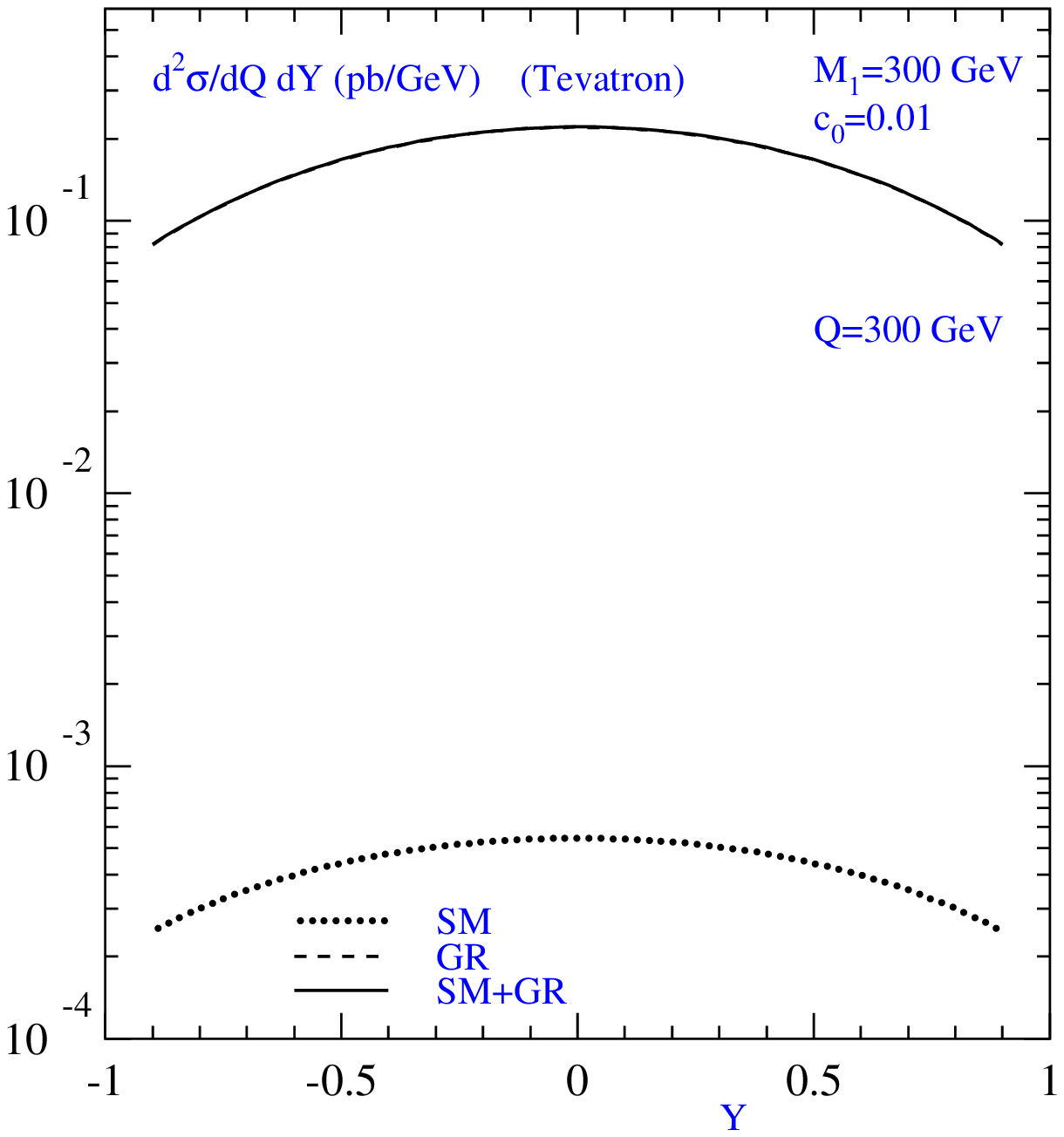,width=15cm,height=16cm,angle=0}}
\vspace{5mm}
\centerline{\bf Fig.~4a}
\end{figure}

\eject

\begin{figure}[htb]
\vspace{1mm}
\centerline{\epsfig{file=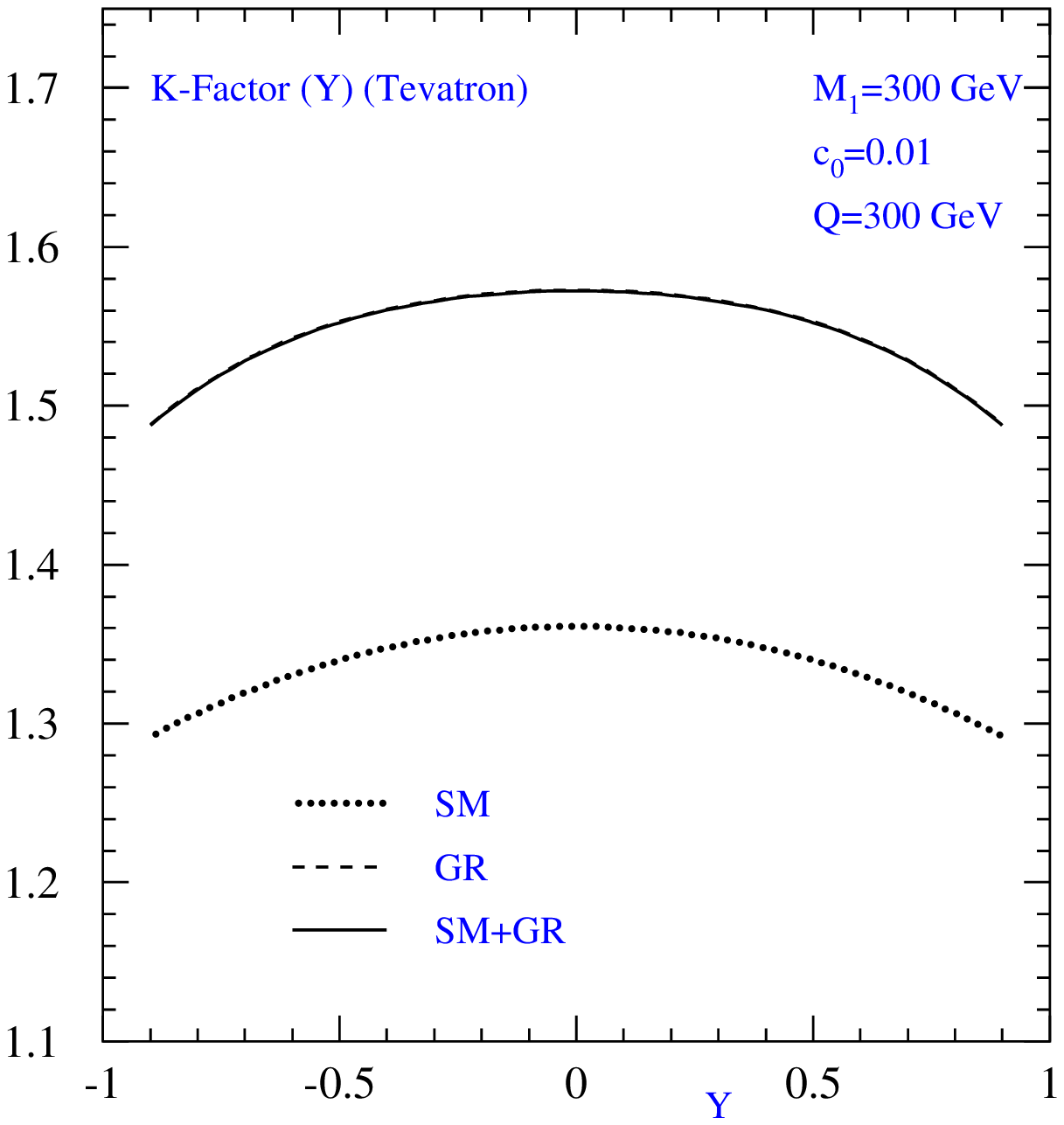,width=15cm,height=16cm,angle=0}}
\vspace{5mm}
\centerline{\bf Fig.~4b}
\end{figure}

\eject

\begin{figure}[htb]
\vspace{1mm}
\centerline{\epsfig{file=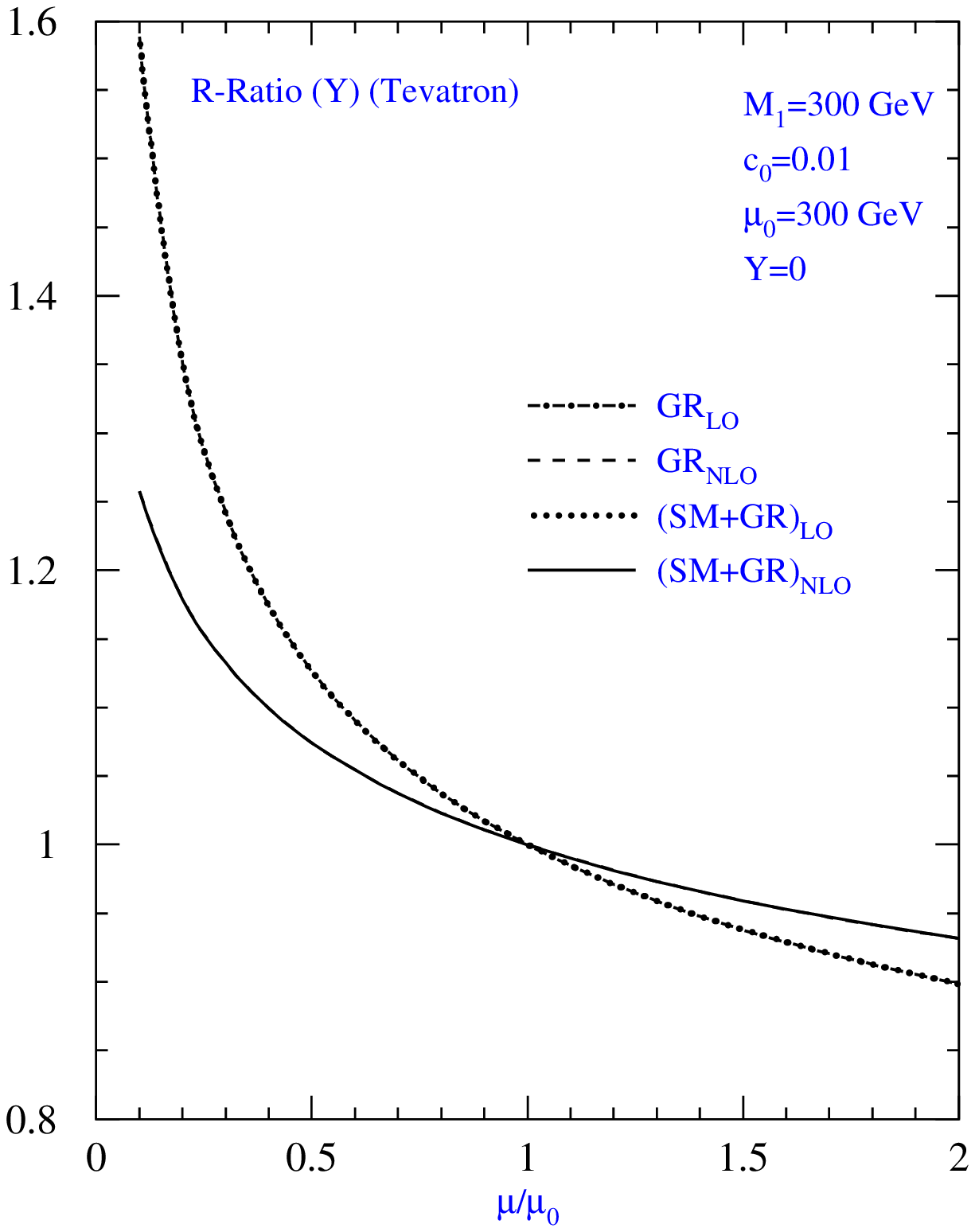,width=15cm,height=16cm,angle=0}}
\vspace{5mm}
\centerline{\bf Fig.~4c}
\end{figure}


 
\end{document}